\documentclass[11pt,letterpaper]{article}
\usepackage[utf8]{inputenc}
\usepackage{ dsfont }

\pdfoutput=1
\usepackage{color}
\definecolor{darkblue}{rgb}{0.1,0.1,.7}
\usepackage[colorlinks, linkcolor=darkblue, citecolor=darkblue, urlcolor=darkblue, linktocpage]{hyperref}
\usepackage[]{amsmath}
\usepackage[]{graphicx}
\usepackage[]{latexsym}
\usepackage[utf8]{inputenc}
\usepackage{slashed,graphicx,color,amsmath,amssymb}
\usepackage[mathscr]{eucal}
\usepackage{mathrsfs}
\usepackage[margin=10pt,font=small,labelfont=bf]{caption}
\usepackage{amscd}
\usepackage{bm}
\usepackage{xcolor}
\usepackage{upgreek}
\usepackage[square, comma, sort&compress,numbers]{natbib}
\usepackage[all,cmtip]{xy}
\usepackage[margin=1.7in]{geometry}
\usepackage{cleveref}
\usepackage[color=cyan!30!white,linecolor=red,textsize=footnotesize]{todonotes}
\usepackage{soul}
\numberwithin{equation}{section}
\usepackage{framed} %%%(For shading)%%%
\definecolor{shadecolor}{rgb}{0.9,0.9,0.95}

\hypersetup{
	pdfstartview={FitH},    % fits the width of the page to the window
	colorlinks=true,       % false: boxed links; true: colored links
	linkcolor=blue,          % color of internal links (change box color with linkbordercolor)
	citecolor=blue!59!black! ,        % color of links to bibliography
	filecolor=magenta,      % color of file links
	urlcolor=blue           % color of external links
}

\usepackage[utf8]{inputenc}
\usepackage{tikz}
\usetikzlibrary{shapes.misc}
\usetikzlibrary{decorations.markings}
\tikzset{cross/.style={cross out, draw=black, ultra thick, minimum size=2*(#1-\pgflinewidth), inner sep=0pt, outer sep=0pt},
cross/.default={5pt}}
\usetikzlibrary{decorations.pathmorphing}

\tikzset{snake it/.style={decorate, decoration=snake}}

\usetikzlibrary{arrows}
\usetikzlibrary{decorations.markings}
\tikzset{
  big arrow/.style={
    decoration={markings,mark=at position 1 with {\arrow[scale=2.5,#1]{>}}},
    postaction={decorate},
    shorten >=0.4pt},
  big arrow/.default=blue}
  \tikzset{
  double arrow/.style={
    decoration={markings,mark=at position 1 with {\arrow[scale=2.5,#1]{>>}}},
    postaction={decorate},
    shorten >=0.4pt},
  big arrow/.default=blue}
  
\usetikzlibrary{calc} 
\def\be{\begin{equation}}
\def\ee{\end{equation}}

\def\l1{{{1-loop}}}

\def\n1{\Bigg|_{n=1}}

\def\n{{(n)}}

\def\beq{\begin{equation}}
\def\eeq{\end{equation}}

\def\bea{\begin{eqnarray}}
\def\eea{\end{eqnarray}}

\def\l1{{\text{1-loop}}}

\def\n1{\Bigg|_{n=1}}

\def\n{{(n)}}

\def\be{\begin{equation}}
\def\ee{\end{equation}}
\def\bal{\begin{array}{l}}
\def\ba#1{\begin{array}{#1}}  %% e.g. \ba{cc}
	\def\ea{\end{array}}
\def\bea{\begin{eqnarray}}
\def\eea{\end{eqnarray}}
\def\beas{\begin{eqnarray*}}
	\def\eeas{\end{eqnarray*}}

\def\bit{\begin{item}}
	\def\eit{\end{item}}
\def\benu{\begin{enumerate}}
	\def\eenu{\end{enumerate}}

\DeclareFontFamily{U}{wncy}{}
\DeclareFontShape{U}{wncy}{m}{n}{<->wncyr10}{}
\DeclareSymbolFont{mcy}{U}{wncy}{m}{n}
\DeclareMathSymbol{\Sha}{\mathord}{mcy}{"58}

 \makeatletter
 \g@addto@macro\bfseries{\boldmath}
 \makeatother
\makeatletter
\if@todonotes@disabled

\else

\fi
\makeatother
\begin{document}

%highlight color
\definecolor{tinge}{RGB}{255, 244, 195}
\sethlcolor{tinge}
\setstcolor{red}

\vspace{.2in} {\Large
\begin{center}
{\LARGE   Positivity of the Veneziano Amplitude in $D=4$} %(Unitarity from Partial Wave Expansion of Veneziano Amplitude)}
\end{center}}
\vspace{.2in}
\begin{center}
{Pronobesh Maity\footnote{Email: \href{mailto:pronobesh.maity@icts.res.in}{pronobesh.maity@icts.res.in}}}
\\
\vspace{.3in}
\small{
\textit{International Centre for Theoretical Sciences,\\
	Shivakote, Hesaraghatta Hobli, Bengaluru North 560 089, India.}
} \vspace{0cm}
%\\ \vspace{.1cm}
%\begingroup\ttfamily\small
%emails\par
%\endgroup

%\vspace{.1in}

\end{center}

\vspace{.5in}

\begin{abstract}
\normalsize

The Veneziano amplitude was put forward as a solution to the axioms of the S-matrix bootstrap. However, unitarity, reflected in the positivity of the coefficients in the Gegenbauer expansion of the amplitude is not obvious. In this note we compute the generating function of these coefficients in terms of the Appell hypergeometric function. We use this to read off an exact form of this coefficient on the leading Regge trajectory in $D=4$. We find that it decays with the spin but always remains positive. Since for large spin these coefficients are expected to be smaller than those on the subleading trajectories, our result indicates the positivity of the full Veneziano amplitude in $D=4$.
\end{abstract}

\vskip 1cm \hspace{0.7cm}
\thispagestyle{empty}
\newpage

\noindent\rule{\textwidth}{.1pt}\vspace{-1.2cm}
\begingroup
\hypersetup{linkcolor=black}
\tableofcontents
\endgroup
\noindent\rule{\textwidth}{.2pt}

\section{Introduction}
\setcounter{page}{1}
Bootstrapping theories with massive particles of arbitrary spin is an old theme of theoretical physics before the advent of quantum chromodynamics. It was motivated by the proliferation of massive higher spin resonances of strongly interacting particles in hadronic physics during 1960s. Arguing for a good high energy behaviour of the tree-level scattering amplitude $A(s,t)$ naturally leads to an infinite sum with an infinite number of exchanged resonances in the following Gegenbauer expansion of $A(s,t)$:
 \begin{equation}\label{partial0}
       A(s,t)=\sum_{n,l}a_{n,l} \,\frac{C_l^{(\alpha)}\left( 1+\frac{2t}{m_{n,l}^2-4m_0^2}\right)}{s-m_{n,l}^2},
\end{equation}
where $C_l^{(\alpha)}$ is the Gegenbauer polynomial and $\alpha=(D-3)/2$. It was no longer clear if we would need to add contributions from t-channel poles separately to the full amplitude. In fact, \cite{Dolen:1967jr} advocated such an approximate equality between s- and t-channels with the help of experimental data. Next in 1968, Veneziano proposed \cite{Veneziano:1968yb} an explicit form of the scattering amplitude
\begin{equation}\label{Veneziano_Model}
    \mathcal{M}(s,t)=\frac{\Gamma(-\alpha's-\alpha_0)\Gamma(-\alpha't-\alpha_0)}{\Gamma(-\alpha's-\alpha't-2\alpha_0)},
\end{equation}
as a solution to the bootstrap axioms, where $\alpha'$ and $\alpha_0$ are known as the Regge slope and Regge intercept respectively. But unitarity, which requires $a_{n,l}\geq 0$, is far from being obvious \cite{Green:1987sp,Arkani-Hamed:2020blm}, see however \cite{Nakazawa}. Subsequent spectacular developments unraveled the origin of Veneziano amplitude in strings and unitarity was then implied \textit{indirectly} from the no-ghost theorem in string theory. Thus we expect for (\ref{Veneziano_Model})
\begin{equation}\label{positivity}
    a_{n,l}\geq 0\quad \text{for} \quad D\leq 26.
\end{equation}
In this note, we will find an explicit form [see eq. (\ref{final})] for this coefficient on the leading Regge trajectory i.e for $a_{n,n+1}$, in $D=4$ space-time dimensions, which falls off exponentially for asymptotic values of the spin but always remains positive. Further all other coefficients on the subleading trajectories are expected (though we couldn't prove it rigorously here) to be larger than the leading one for large spins and thus our result indicates positivity of the Veneziano amplitude in $D=4$.    
\par 
Later the Veneziano amplitude was abandoned as a candidate for the strong interactions due to its soft exponential fall-off in the hard scattering regime compared to the power law fall off observed for strong interactions understood in parton model terms and supported by experimental data. In fact, the huge degeneracy of spins at a resonance of given mass in (\ref{Veneziano_Model}) is absent for QCD. Hence, in the light of revival of $S$-matrix bootstrap program \cite{Caron-Huot:2016icg, Paulos:2016fap, Correia:2020xtr}, it might well be important to explicate (\ref{positivity}) which can help to constructively modify the Veneziano amplitude in $D=4$ and remove the unwanted degeneracy to come closer to the QCD amplitude. This was the original motivation \cite{Komargodski} for this work and we expect that the techniques of this note will be helpful in the future, in this regard.

\section{Bootstrap axioms and Gegenbauer expansion}
We consider a $2\to 2$ scattering of identical lightest scalar glueballs of mass $m_0$. From simple large $N$ counting, such an amplitude goes as $\frac{1}{N^2}$ indicating that the particles almost fly past each other in the large $N$ limit. We can define the usual Mandlestam invariants
\begin{equation}
    s=(k_1+k_2)^2,\quad t=(k_1-k_3)^2,\quad u=(k_1-k_4)^2
\end{equation}
The amplitude will generally involve contributions from all three channels, where $s$ and $u$ (but not $t$) channel diagrams have poles in $s$ at the locations of the resonances. For the sake of simplicity we can forget the u-channel resonances, we can always arrange such a process with no contribution from the $u$-channel with the correct quantum numbers by judiciously choosing a scattering of non-identical particles. 
\par
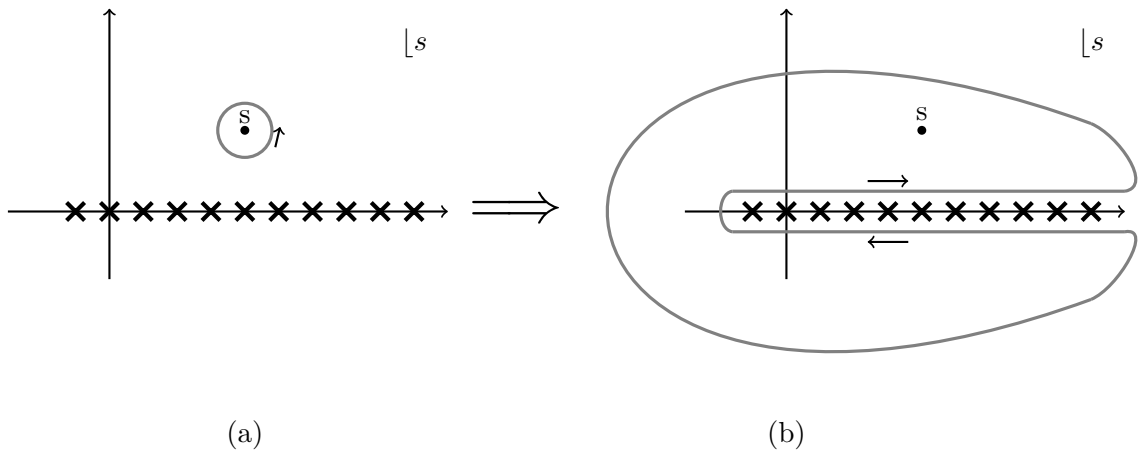
\begin{figure}[!htb]
    \centering
\begin{tikzpicture}[scale=0.9]
%\hspace*{-1cm}% This shifts the whole figure horizontally !!!
\clip (-2,-4)--(17,-4)--(17,3)--(-2,3)--(-2,3);
\draw[thick,->] (0,-1)--(0,3);
\draw[thick,->] (-1.5,0)--(5,0);
\draw[thick, ->] (2.45,0.95)--(2.52,1.25);
\draw (-0.5,0) node[cross] {};
\draw (0,0) node[cross] {};
\draw (0.5,0) node[cross] {};
\draw (1,0) node[cross] {};
\draw (1.5,0) node[cross] {};
\draw (2,0) node[cross] {};
\draw (2.5,0) node[cross] {};
\draw (3,0) node[cross] {};
\draw (3.5,0) node[cross] {};
\draw (4,0) node[cross] {};
\draw (4.5,0) node[cross] {};

\draw[very thick, gray] (2,1.2) circle (0.4cm);
\draw[thick, fill] (2,1.2) circle (0.05cm);

\node at (6,0) {\scalebox{2}{$\Longrightarrow$}};

\draw[thick,->] (0+10,-1)--(0+10,3);
\draw[thick,->] (-1.5+10,0)--(5+10,0);
\draw (-0.5+10,0) node[cross] {};
\draw (0+10,0) node[cross] {};
\draw (0.5+10,0) node[cross] {};
\draw (1+10,0) node[cross] {};
\draw (1.5+10,0) node[cross] {};
\draw (2+10,0) node[cross] {};
\draw (2.5+10,0) node[cross] {};
\draw (3+10,0) node[cross] {};
\draw (3.5+10,0) node[cross] {};
\draw (4+10,0) node[cross] {};
\draw (4.5+10,0) node[cross] {};

\draw[thick, fill] (2+10,1.2) circle (0.05cm);
\node at (2,1.4) {s};
\node at (2+10,1.45) {s};
\node at (4.5,2.5) {$\lfloor s$};
\node at (4.5+10,2.5) {$\lfloor s$};
\node at (2,-3.3) {(a)};
\node at (2+8,-3.3) {(b)};
\draw[very thick, gray] (-0.8+10,0.3)--(15,0.3) (-0.8+10,-0.3)--(15,-0.3);
\draw[thick,->] (11.2,0.45)--(11.8,0.45);
\draw[thick,->] (11.8,-0.45)--(11.2,-0.45);
%\draw[very thick, cyan] (15,0.3) to[out=140, in=-140, looseness=55] (15,-0.3);
\draw[very thick, gray] (15,0.3) to[out=10,in=-23] (14.5,1.3);
\draw[very thick, gray] (15,-0.3) to[out=10,in=23] (14.5,-1.3);
\draw[very thick, gray] (14.5,1.3) to[out=160,in=-160, looseness=10] (14.5,-1.3);
\draw[very thick, gray] (-0.8+10,0.3) to[out=190,in=170] (-0.8+10,-0.3);
\end{tikzpicture}    
    \caption{We can deform the contour in (a) into (b) for computing the scattering amplitude in (\ref{contour_expression}). The poles have been drawn in accordance with the Veneziano amplitude (\ref{Veneziano}).}
    \label{fig:boundedness_figure}
\end{figure}

 
Before the developments of QCD, people tried to extract the physics of hadrons and glueballs by imposing a set of natural postulates on the scattering amplitudes $A(s,t)$, outlined below: 
\begin{enumerate}
    \item \textbf{Crossing symmetry:}
    \begin{equation}\label{crossing}
        A(s,t)=A(t,s).
    \end{equation}
    This is manifest for scattering of identical particles in the description in terms of Feynman diagrams, see \cite{Mizera:2021fap} for recent developments on an interesting physical interpretation behind it.

    \item \textbf{Analyticity:} $A(s,t)$ is a meromorphic function in $s$ with poles only on the real axis $s=m_{n,l}^2$, corresponding to the spectrum of the theory (See \cite{Meyer:2004jc} for some fascinating computation of glueball masses for the first few spins from lattice calculations)
    \begin{equation}\label{Imaginary}
    \begin{split}
        \text{Im}[A(s,t)]=\sum_{n,l} f_{n,l}^2\,\delta(s-m_{n,l}^2)\,C_l^{(\alpha)}\left( 1+\frac{2t}{m_{n,l}^2-4m_0^2}\right).
    \end{split}
\end{equation}
    where $C^{(\alpha)}_l(x)$ is the Gegenbauer polynomial which can be expressed in terms of Gaussian hypergeometric (finite) series:
\begin{equation}
    C^{(\alpha)}_l(x)=\frac{(2\alpha)_l}{l!}{}_2F_{1}\left(-l,2\alpha+l; \alpha+\frac{1}{2};\frac{1-x}{2} \right)
\end{equation}
and 
\begin{equation}\label{alpha}
    \alpha=\frac{D-3}{2}.
\end{equation}
In particular for $D=4$,
\begin{equation}
    C_l^{\frac{1}{2}}(z)=P_l(z)
\end{equation}
where $P_l(z)$ is the standard Legendre polynomial. 
    This analyticity behaviour (\ref{Imaginary}) is a consequence of casuality, which states that two regions with a space-like separation don't influence each other.

    \item \textbf{Unitarity:}
    In some sense, unitarity is automatically satisfied, since the amplitude goes as $\mathcal{O}\left(\frac{1}{N^2}\right)$ and hence always remains much less than $1$ in the large $N$ limit. Still it imposes the following non-trivial constraints since couplings $\{f_{n,l}\}$ at both junctions (see figure \ref{fig:unitarity}) are the same for any exchanged resonance in the scattering of identical particles:
    $$f_{n,l}^2>0.$$
    so that the sum in (\ref{Imaginary}) is positive-definite. 

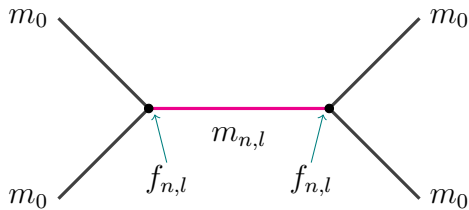
\begin{figure}[!htb]
    \centering
  \begin{tikzpicture}[scale = 0.8]
\draw[very thick, magenta] (0,0)--(3,0);
\draw[very thick, darkgray] (-1.5,1.5)--(0,0) (-1.5,-1.5)--(0,0) (3,0)--(4.5,1.5) (3,0)--(4.5,-1.5);
\filldraw[fill=black] (0,0) circle (2pt) (3,0) circle (2pt);
\node at (1.5,-0.5) {\large $m_{n,l}$};

\node at (-2,1.5) {\large $m_{0}$};
\node at (-2,-1.5) {\large $m_{0}$};
\node at (5,1.5) {\large $m_{0}$};
\node at (5,-1.5) {\large $m_{0}$};

\draw[->,teal] (0.3,-0.9)--(0.1,-0.1);
\node at (0.3,-1.2) {\large $f_{n,l}$};

\draw[->, teal] (2.7,-0.9)--(2.9,-0.1);
\node at (2.7,-1.2) {\large $f_{n,l}$};

\end{tikzpicture}
    \caption{Scattering process with exchanged resonance of mass $m_{n,l}$.}
    \label{fig:unitarity}
\end{figure}

    \item \textbf{Boundedness:} There exists a $t_0$ and a particle of spin $L$ in the spectrum such that
    \begin{equation}\label{boundedness}
        \lim_{|s|\to \infty} s^{-L}A(s,t_0)=0.
    \end{equation}
    From a modern perspective \cite{Camanho:2014apa}, it can be motivated as an implication of causality for a theory with exchange of a massive higher spin particle of spin $L>2$. In the hard scattering regime, scattering amplitude of strong interaction falls off according to a power law and so (\ref{boundedness}) is naturally satisfied for some large negative $t$ in that case. 
\end{enumerate}
We will now explore some immediate consequences of these four axioms on $A(s,t)$: Given the boundedness condition for some region of $t$ for $L=0$ [\textcolor{gray}{axiom 4}], we can exploit the Cauchy integral formula to write
\begin{equation}\label{contour_expression}
    \begin{split}
        A(s,t)=\oint_{C_s}ds'\,\frac{A(s',t)}{s'-s}
    \end{split}
\end{equation}
where the contour $C_s$ doesn't enclose any of the poles of $A(s,t)$ as shown in the figure \ref{fig:boundedness_figure} (a). We can always deform the contour as in fig \ref{fig:boundedness_figure} (b) to rewrite it as
\begin{equation}\label{unsubtracted_dispersion_relation}
    \begin{split}
        A(s,t)=\int_{m_s^2-|\epsilon|}^{\infty} ds' \, \frac{\text{Im}[A(s',t)]}{s'-s}.
    \end{split}
\end{equation}
which is known as the unsubtracted dispersion relation. Plugging (\ref{Imaginary}) [\textcolor{gray}{axiom 2}] into the above relation (\ref{unsubtracted_dispersion_relation}), we immediately get
   \begin{equation}\label{partial}
        A(s,t)=\sum_{n,l}f_{n,l}^2 \,\frac{C_l^{(\alpha)}\left( 1+\frac{2t}{m_{n,l}^2-4m_0^2}\right)}{s-m_{n,l}^2}.
    \end{equation}
Unitarity [\textcolor{gray}{axiom 3}] says $f_{n,l}^2\geq 0$. Crossing symmetry [\textcolor{gray}{axiom 1}] implies  $A(s,t)$ in (\ref{partial}) must have poles in $t$ at the same positions $\{m_{n,l}^2\}$ as for the poles in $s$. It automatically suggests \cite{Caron-Huot:2016icg} that the sum in (\ref{partial}) is an infinite sum, i.e with infinitely many resonances. In fact, there shouldn't be any upper limit on the spin $l$ of exchanged resonance, since otherwise if there exists such an $l_{\text{max}}$, then $\left(\frac{\partial}{\partial t}\right)^{l_{\text{max}}+1}A(s,t)=0$, contradicting with (\ref{crossing}).

% Note that we have ignored the contributions from the u-channel poles residing in the negative real s-axis.

\section{Coefficient of the Veneziano amplitude}
It is a difficult mathematical problem to find an explicit solution to those four bootstrap axioms. The only known (upto tiny dressings for Heterotic strings) solution proposed by Veneziano \cite{Veneziano:1968yb} is the famous Veneziano amplitude 
\begin{equation}\label{Veneziano}
    \mathcal{M}(s,t)=\frac{\Gamma(-s-1)\Gamma(-t-1)}{\Gamma(-s-t-2)}.
\end{equation}
It describes the scattering amplitude of four identical tachyons of mass $m_0^2=-1$. 
\par 
Clearly the amplitude (\ref{Veneziano}) is crossing symmetric with poles in $s$ (and $t$) at $s=n$ for $n=-1,0,1,2,...$ Boundedness follows from the asymptotic behaviour of Gamma functions for large $s$ and fixed $t$
\begin{equation}
    \mathcal{M}(s,t)\sim \Gamma(-t-1)s^{t+1}.
\end{equation}
Thus for $t<-1$, (\ref{boundedness}) is satisfied. Now to check unitarity, we first calculate the residue at the poles in $s$ from (\ref{Veneziano})
\begin{equation}\label{first}
    \text{Res}_{s=n} \mathcal{M}(s,t) = \frac{(-1)^n}{(n+1)!}\frac{\Gamma(-t-1)}{\Gamma(-t-2-n)}= -\frac{1}{(n+1)!}(t+2)_{n+1}.
\end{equation}
From the Gegenbauer  expansion in general dimensions,
\begin{equation}\label{second}
   - \text{Res}_{s=n}\mathcal{M}(s,t)=\sum_{l=0}^{n+1} a_{n,l} C_l^{(\alpha)} \left( 1+\frac{2t}{n+4}\right).
\end{equation}
We note that Veneziano amplitude has huge degeneracy in the spin with particles of spin $l=0$ to $l=(n+1)$ for the same mass $m_{n,l}^2=n$. This degeneracy is drastically different from what is observed for QCD. Equating two expressions in (\ref{first}) and (\ref{second}), we get the following
\begin{equation}
    \sum_{l=0}^{n+1} a_{n,l} C_l^{(\alpha)} \left( 1+\frac{2t}{n+4}\right)=\frac{1}{(n+1)!}(t+2)_{n+1}.
\end{equation}
We can now use the orthogonality relation of Gegenbauer polynomials 
\begin{equation}
    \int_{-1}^{+1} dx\,C_{l}^{(\alpha)} (x)\, C_{l'}^{(\alpha)}(x)\, (1-x^2)^{\alpha-\frac{1}{2}} = 2\underbrace{ \frac{\pi\Gamma(l+2\alpha)}{2^{2\alpha}l! (l+\alpha) [\Gamma(\alpha)]^2}}_{\Lambda(l,\alpha)} \delta_{ll'},
\end{equation}
to extract out $a_{n,l}$ as
\begin{equation}\label{a_{n,l}}
    a_{n,l}=\frac{1}{\Lambda(l,\alpha)}\frac{1}{(n+4)(n+1)!} \left[\frac{4}{(n+4)^2}\right]^{\alpha-\frac{1}{2}}\int_{0}^{n+4} dt\, C_{l}^{(\alpha)}\left(1-\frac{2t}{n+4}\right) (-t+2)_{n+1} [t(n+4-t)]^{\alpha-\frac{1}{2}}.
\end{equation}

%\begin{equation}
 %   a_{n,l}= \frac{2l+1}{(n+4)(n+1)!} \int_{0}^{n+4} dt P_l \left(1-\frac{2t}{n+4}\right) (-t+2)_{n+1}
%\end{equation}

Note that within the integration range, Legendre polynomial changes sign (see figure \ref{fig:Legendre}) and a priori it is not obvious that $a_{n,l}\geq 0$. In fact, we can show that (nearly) half of such coefficients vanish due to the following result:
\begin{figure}[!htb]
    \centering
    \includegraphics[width=0.6
    \textwidth]{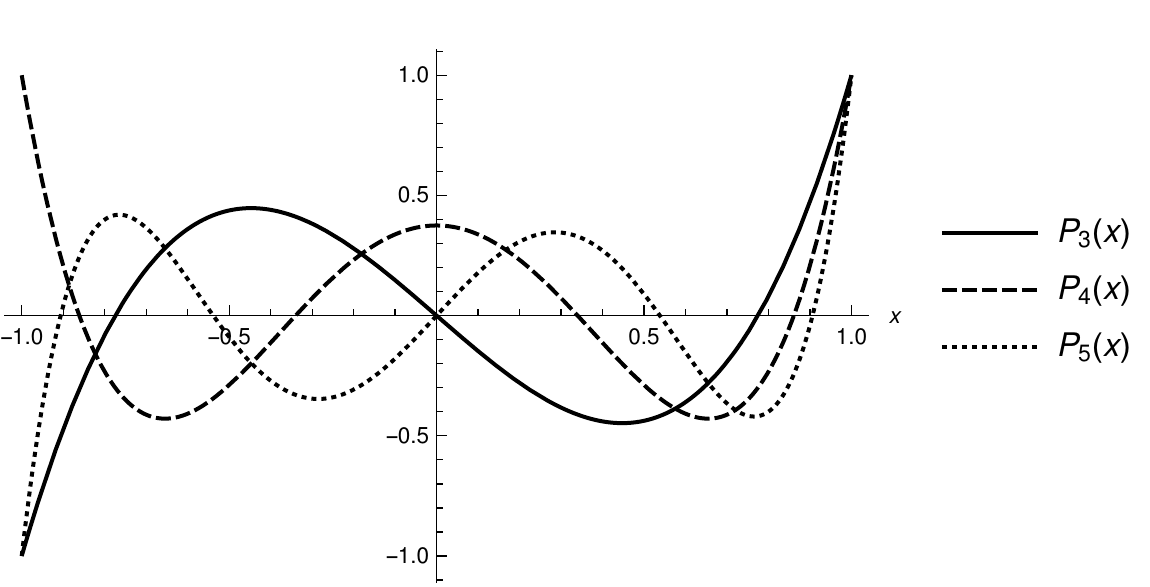}
    \caption{Few Legendre polynomials and their oscillating behaviours}
   \label{fig:Legendre}
\end{figure}

\textbf{Claim:} $a_{n,l}=0$ when $(n+l)$ is even.
\vspace{2mm}

\begin{figure}[!htb]
    \centering
    \includegraphics[width=0.5
    \textwidth]{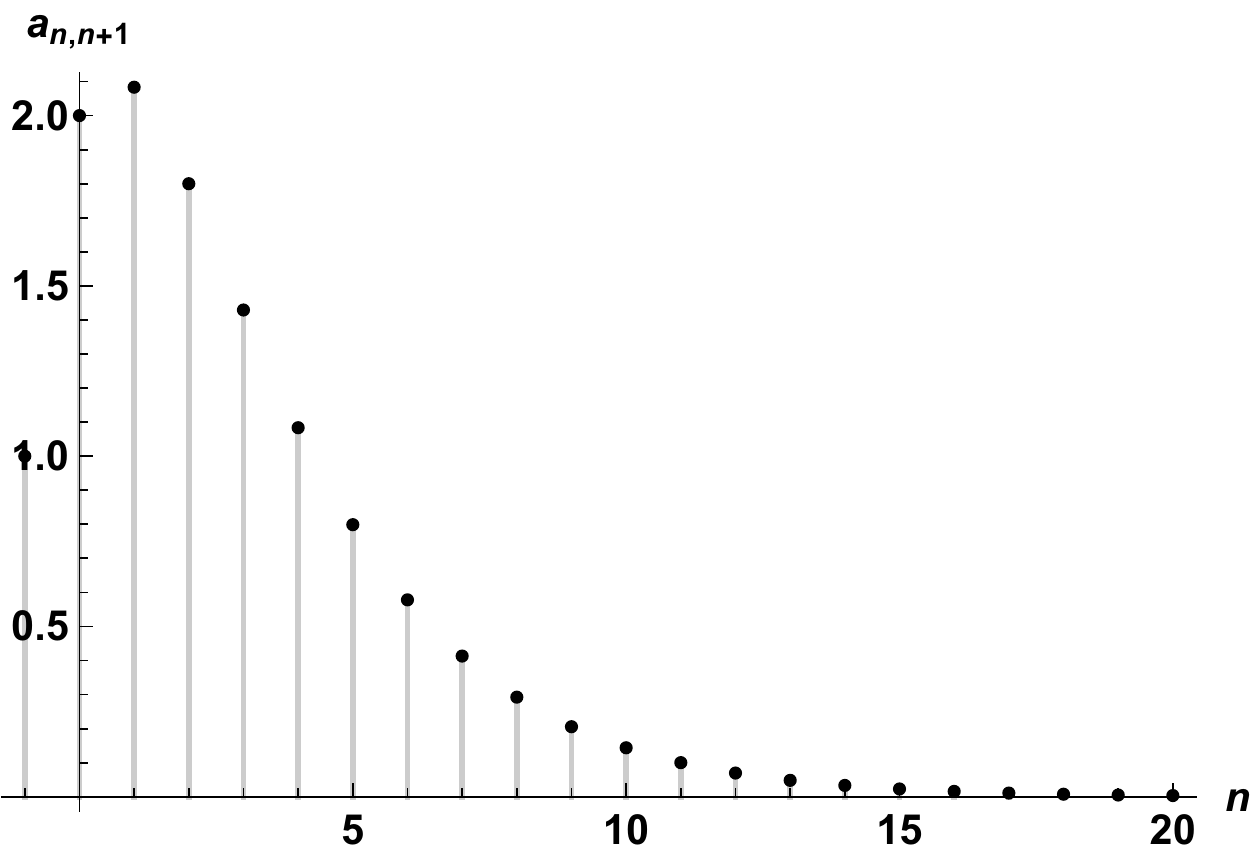}
    \caption{Coefficients on the leading Regge trajectory in $D=4$.}
   \label{fig:Observation1b}
\end{figure}

\textbf{Proof:}
Substituting $t=n+4-t'$ in (\ref{a_{n,l}}), the integral can be written as,
\begin{equation}
\begin{split}
    \int_{0}^{n+4} dt \, C_l^{(\alpha)} \left(1-\frac{2t}{n+4}\right)&(-t+2)_{n+1} [t(n+4-t)]^{\alpha-\frac{1}{2}}\\&= (-)^{n+l+1}\,\int_{0}^{n+4} dt' C_l^{(\alpha)} \left(1-\frac{2t'}{n+4}\right) (-t'+2)_{n+1} [t'(n+4-t')]^{\alpha-\frac{1}{2}},
    \end{split}
\end{equation}
where we used $(-x)_{m}=(-1)^{m}(x-m+1)_{m}$ and $C_{l}^{(\alpha)}(-x)=(-1)^lC_l^{(\alpha)}(x)$. Thus, 
\begin{equation}
    [1+(-1)^{n+l}]a_{n,l}=0.
\end{equation}
This establishes the claim [QED]. Note that the coefficients $a_{n,n+1}$ on the leading Regge trajectories are always non-zero. We have plotted $a_{n,n+1}$ in figure \ref{fig:Observation1b} for first few values of $n$ from (\ref{a_{n,l}}) using Mathematica. Clearly they decrease rapidly with $n$, but seem to remain positive as far as we can check numerically. 

\section{Coefficient on the leading Regge trajectory in $D=4$}

\begin{figure}[!htb]
    \centering
    \includegraphics[width=0.6
    \textwidth]{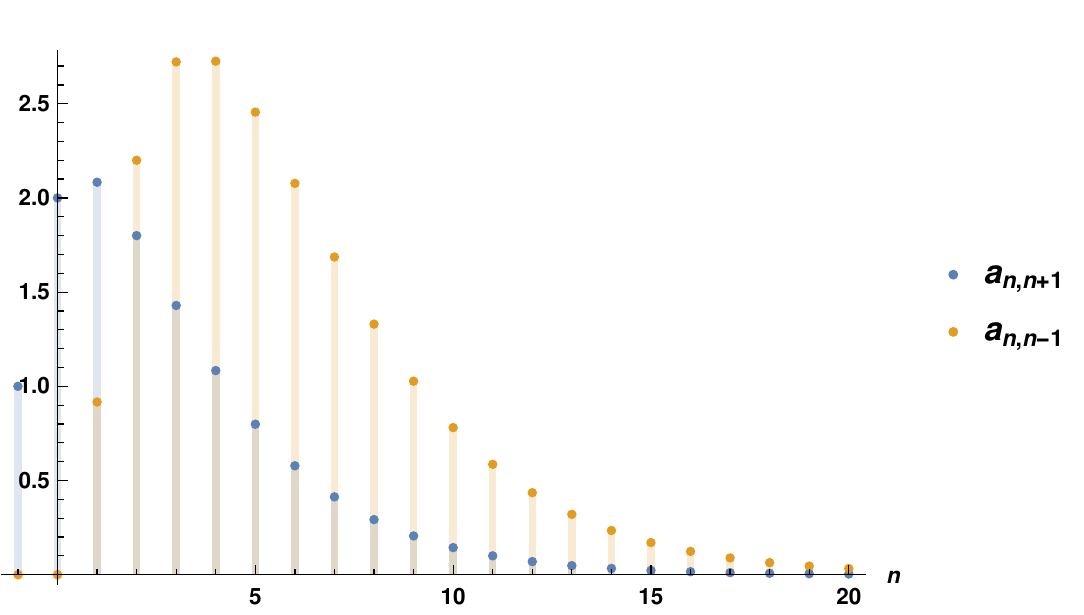}
    \caption{Coefficients on the leading and sub-leading Regge trajectories in $D=4$.}
   \label{fig:Leading_Subleading}
\end{figure}

In this note we will only be able to calculate an explicit form of the coefficient on the leading Regge trajectory in $D=4$. In fact we can check (though not proved yet) that all other coefficients on the subleading trajectories are larger than the leading one for asymptotic values of the spin (see figure \ref{fig:Leading_Subleading} for coefficients $a_{n,n+1}$ and $a_{n,n-1}$, remember $a_{n,n}=0$). It is thus expected that positivity of $a_{n,n+1}$ would imply positivity for all others.

The expression of $a_{n,l}$ in (\ref{a_{n,l}}) simplifies for $D=4$ as
\begin{equation}\label{inD=4}
    a_{n,l}= \frac{2l+1}{(n+4)(n+1)!} \int_{0}^{n+4} dt \, P_l \left(1-\frac{2t}{n+4}\right) (-t+2)_{n+1}.
\end{equation}
First we construct a set of generating functions\footnote{Here h has been analytically continued to any positive real number (both $>1$ and $<1$).} of $a_{n,l}$ for each value of $n$ as
\begin{equation}\label{Generating_Function_1}
   G_{n}(h)=\sum_{j=0}^{\infty}\frac{1}{2j+1}\frac{a_{n,j}}{h^{j+1}}= \frac{1}{(n+4)(n+1)!} \int_{0}^{n+4} dt\,\frac{(-t+2)_{n+1}}{[(h-1)^2+\frac{4ht}{n+4}]^{1/2}},
\end{equation}
where we have exploited the standard result
\begin{equation}
    \sum_{l=0}^{\infty}P_l(x)\,t^{-l-1}=\frac{1}{\sqrt{1-2xt+t^2}}.
\end{equation}
We can then use the following representation for the Pochhammer symbol,
\begin{equation}\label{series_expansion}
    (-t+2)_{n+1}=\sum_{k=0}^{n+1} (-1)^{n+1-k}\,s(n+1,k)\,(-t+2)^k,
\end{equation}
where $s(m,p)$ is the Stirling number of the first kind, which is defined as $(-1)^{m-p}$ times the number of permutations of $\{1,2,...,m\}$ with exactly p cycles. Thus,
\begin{equation}\label{Generating_Function_2}
    G_n(h)=\frac{1}{(n+4)(n+1)!}  \sum_{k=0}^{n+1} (-1)^{n+1-k}\,s(n+1,k)\, g_{n,k}(h),
\end{equation}
with the `pseudo-generating function' $g_{n,k}(h)$ defined by
\begin{equation}
    g_{n,k}(h)= \int_{0}^{n+4} dt\,\frac{(-t+2)^k}{[(h-1)^2+\frac{4ht}{n+4}]^{1/2}},
\end{equation}
which can be explicitly evaluated as,
\begin{equation}\label{g[n,k]1}
\begin{split}
    g_{n,k}(h)=\frac{1}{2^{2k+1}}\, \frac{(n+4)^{k+1}}{h^{k+1}} \left[(h-1)^2+\frac{8h}{n+4}\right]^k \Big[& (h+1) {}_2F_{1}\left( \frac{1}{2},-k;\frac{3}{2};\frac{(h+1)^2}{(h-1)^2+\frac{8h}{n+4}} \right)\\&-(h-1){}_2F_{1}\left( \frac{1}{2},-k;\frac{3}{2};\frac{(h-1)^2}{(h-1)^2+\frac{8h}{n+4}}\right)\Big].
    \end{split}
\end{equation}
More compactly, we can represent it in terms of Appell hypergeometric function of two variales $\text{Appell}F_{1}(a;b_1,b_2;c;x,y)$:
\begin{equation}\label{another_representation}
     g_{n,k}(h)=2^k\frac{n+4}{h-1}\,\text{Appell} F_{1}\left(1;-k,\frac{1}{2};2;\frac{n+4}{2},-\frac{4h}{(h-1)^2}\right),
\end{equation}
whose precise form can be found in (\ref{Appell_expansion}). 
\par 
To analyse (\ref{g[n,k]1}) further, we can simplify it using the definition of ${}_2F_{1}$
\begin{equation}\label{2F1}
    {}_2F_{1}(a,b;c;z)=\sum_{p=0}^{\infty}\frac{(a)_p(b)_p}{(c)_p\,p!}z^p,
\end{equation}
to get
\begin{equation}\label{g[n,k]2}
    g_{n,k}(h)=\frac{1}{2^{2k+1}}\, \frac{(n+4)^{k+1}}{h^{k+1}} \sum_{p=0}^{k} \frac{(-k)_p}{(2p+1)p!} \left[(h-1)^2+\frac{8h}{n+4}\right]^{k-p}\sum_{m=0}^{2p+1} \binom{2p+1}{m} h^m (1+(-1)^m). 
\end{equation}
Note that the $p$-sum in (\ref{2F1}) terminates in the above expression (\ref{g[n,k]2}) of $g_{n,k}(h)$ since one of the arguments of ${}_2F_1$ is a negative integer: $b=-k$ and $(-k)_{p}=0\;\text{for}\;p>k$.
\par 
Now $g_{n,k}(h)$ in (\ref{g[n,k]2}) has $(k+1)$ terms with $h^{-1}\cdots h^{-k-1}$. We can find the coefficient of $h^{-k-1}$ from that expansion as
\begin{equation}
  d_{n,k} = \frac{(n+4)^{k+1}}{2^{2k+1}}\,  \frac{\sqrt{\pi} k!}{\Gamma(k+3/2)}.
\end{equation}
To find the coefficient on the leading Regge trajectory, we need $a_{n,n+1}$, i.e the term with $h^{-n-2}$ in the expression of $G_n(h)$ in (\ref{Generating_Function_1}). From the above analysis of $g_{n,k}(h)$, we can rewrite $h$-expansion of $G_n(h)$ as
\begin{equation}\label{rewrite}
    \underbrace{\sum_{j=0}^{n+1}[...]\,h^{-j-1}}_{G_n(h)}=\sum_{k=0}^{n+1}[...]\,\underbrace{\sum_{q=0}^{k}[...]\,h^{-q-1}}_{g_{n,k}(h)}.
\end{equation}
From this expression (\ref{rewrite}), $h^{-n-2}$ term on the R.H.S corresponds to taking the unique term with $h^{-n-2}$ in $g_{n,n+1}(h)$. Note that for subleading trajectories, we would require $h^{-k-1}$ with $k<n$ and there should be contributions from several terms in the R.H.S of (\ref{rewrite}), as discussed thorougly in appendix \ref{appendixA}.  
\par 
Hence equating coefficients of $h^{-n-2}$ from both sides of (\ref{rewrite}), we get
\begin{equation}
\frac{1}{2n+3}a_{n,n+1} = \frac{1}{(n+4)(n+1)!} s(n+1,n+1) \,d_{n,n+1}.
\end{equation}
This gives the explicit form of $a_{n,n+1}$
\begin{shaded}
\begin{equation}\label{final}
    a_{n,n+1}=\frac{\sqrt{\pi}}{2^{2n+2}}\frac{(n+4)^{n+1}}{\Gamma(n+\frac{3}{2})} ,\; \forall \;n\geq -1.
\end{equation}
\end{shaded}
It reproduces the plot in figure \ref{fig:Observation1b}. Note that for asymptotic values of $n$, it decays exponentially:
\begin{equation}\label{decay}
    a_{n,n+1}\sim \exp[-n\log(4/e)],
\end{equation}
but obviously remains positive.

\section{Discussion}\label{Discussion}
In this note, we found an explicit form for the coefficients in the Gegenbauer expansion of the Veneziano amplitude for the leading Regge trajectory in $D=4$. They decay exponentially for large spin but remain positive. We also mentioned that the coefficients on the subleading trajectories are expected to be larger than those on the leading ones for large spin and thus our result is a strong indication of the positivity of the Veneziano amplitude in $D=4$. 
\par 
As mentioned in the introduction, one long-term motivation of this work was to understand the consistent deformations of the Veneziano amplitude satisfying the Bootstrap axioms to bring it closer to have features of 4d QCD. In fact, some attempts to generalize the Veneziano amplitude have already been advocated in \cite{Coon:1969yw,Fairlie:1994ad}, suggesting the `Coon amplitude' \cite{Caron-Huot:2016icg}. It has the explicit form \cite{Fairlie:1994ad}
\begin{equation}
    A(s,t)\sim \prod_{r=0}^{\infty} \frac{[(\sigma-1)(s-m^2)+1][(\sigma-1)(t-m^2)+1]-\sigma^r}{[(\sigma-1)(s-m^2)+1-\sigma^r][(\sigma-1)(t-m^2)+1-\sigma^r]},
\end{equation}
where at $\sigma=1$ it reduces to the Veneziano amplitude. This generalized amplitude appears to be unitary \cite{Fairlie:1994ad} for $0\leq \sigma \leq 1$ from numerical verification of the positivity of some set of coefficients $\{a_{n,l}\}$. It would be very gratifying if our analytical strategy can be utilized to show this explicitly.  
\par 
It is also interesting to note that the exponential decay of the coefficient in (\ref{decay}) is somewhat similar to the decay of the partial wave amplitudes with angular momentum quantum number in simple quantum mechanical scattering processes \cite{Sakurai:2011zz}. More explicitly, for spherically symmetric scattering potential, the ``scattering amplitude" $f(\Vec{k'},\Vec{k})$ in the wavefuntion 
\begin{equation}
    \langle \Vec{x}|\Psi\rangle \to A\left[ e^{i\Vec{k}\cdot \Vec{x}}+f(\Vec{k'},\Vec{k})\frac{e^{ikr}}{r}\right]\quad r>>1,
\end{equation}
has the familiar partial wave expansion
\begin{equation}
    f(\Vec{k'},\Vec{k})=f(\Vec{k},\theta) =\sum_{l=0}^{\infty} \underbrace{(2l+1) \left[ \frac{e^{2i\delta_l}-1}{2ik}\right]}_{a_l} P_l(\cos{\theta}).
\end{equation}
For small $\delta_l$, $ a_l\sim (2l+1) \frac{\delta_l}{k}$. As a concrete example, consider hard sphere scattering with the potential
\begin{equation} \label{hard_sphere_scattering}
    V(r)=\begin{cases}
    \infty \quad r<R,\\
    0 \quad \quad r>R.
    \end{cases}
\end{equation}
Then in the low-energy limit with $kR<<1$ and large $l$ 
\begin{equation}
    a_l\sim \frac{1}{k} \exp\left[-2l\log\left(\frac{2l}{kRe}\right)\right].
\end{equation}
In our context of the Veneziano amplitude, two incoming strings do feel some ``force" between them though that should not be modelled by hard sphere potential (\ref{hard_sphere_scattering}). In any case, the connection seems worth exploring.

\appendix 

\section*{Acknowledgements}
I would like to thank Zohar Komargodski for useful correspondence that led to some of the comments in section \ref{Discussion}, and Rajesh Gopakumar for very detailed and helpful suggestions on an earlier version of this draft. This work was initiated by the question posed in a fascinating YouTube talk \cite{Komargodski}. It is supported by the Department of Atomic Energy, Government of India, under project no. RTI4001.

\appendix 

\section{General coefficient in $D=4$}\label{appendixA}
In this section, we will express the coefficients on general Regge trajectories in relatively simpler ways than (\ref{inD=4}). To get the general term with $h^{-l-1}$ from both sides of (\ref{rewrite}), we can proceed with the binomial expansion of $g_{n,k}(h)$ in (\ref{g[n,k]1}) to extract the coefficient of $h^{-l-1}$ and then finally use (\ref{Generating_Function_2}) to write the generic coefficient $a_{n,l}$ as
\begin{equation}
\begin{split}
   a_{n,l}=\frac{2l+1}{(n+1)!}&\sum_{k=0}^{n+1}(-)^{n+1-k} s(n+1,k) \frac{1}{2^{2k+1}} (n+4)^{k} \\&\times\sum_{p=0}^{k}\sum_{m=0}^{2p+1}\sum_{r=0}^{k-p}\binom{2p+1}{m}\binom{k-p}{r} \binom{2r}{p+r-l-m} \frac{(-k)_p}{(2p+1)p!} \\& \times \left( \frac{8}{n+4}\right)^{k-p-r} (-1)^{p+r-l-m} (1+(-1)^m).
    \end{split}
\end{equation}
Though looks clumsy, the above Eq. appears algorithmically much simpler than the original expression (\ref{a_{n,l}})
\par 
We can also use another representation (\ref{another_representation}) of $g_{n,k}(h)$, 
\begin{equation}\label{Appell_expansion}
\begin{split}
    g_{n,k}(h)&=2^k\frac{n+4}{h-1}\,\text{Appell} F_{1}\left(1;-k,\frac{1}{2};2;\frac{n+4}{2},-\frac{4h}{(h-1)^2}\right)\\ &= 2^k\frac{n+4}{h-1}\, \sum_{p=0}^{k} \sum_{q=0}^{\infty} \frac{(-k)_p(\frac{1}{2})_q}{p!q!(1+p+q)}\left( \frac{n+4}{2}\right)^p \left(-\frac{4h}{(h-1)^2} \right)^q.
    \end{split}
\end{equation}
We can actually perform the $p$-sum to finally get 
\begin{equation}\label{another}
     g_{n,k}(h)= 2^k\frac{n+4}{h-1} \sum_{q=0}^{k} \frac{(\frac{1}{2})_q}{(q+1)!} {}_2F_{1}\left(-k,1+q;q+2;\frac{n+4}{2} \right) \left(-\frac{4h}{(h-1)^2}\right)^q.
\end{equation}
We can similarly extract the coefficient of $h^{-l-1}$ from (\ref{another}) to write the general coefficient
\begin{equation}
    \begin{split}
        a_{n,l}=\frac{2l+1}{(n+4)(n+1)!}&\sum_{k=0}^{n+1}(-1)^{n+1-k} s(n+1,k) 2^k (n+4) \sum_{q=0}^{l}\frac{(\frac{1}{2})_q}{(q+1)!} {}_2F_{1}\left(-k,1+q;q+2;\frac{n+4}{2}\right)\\& \times (-4)^q \frac{(2q+1)_{l-q}}{(l-q)!}.
    \end{split}
\end{equation}
This has relatively simmpler form with only two sums (instead of 4-sums in the last expression). Also the $l=0$ case, $a_{n,0}$ has only a single sum:
\begin{equation}
    \begin{split}
        a_{n,0}=\frac{1}{(n+4)(n+1)!} \sum_{k=0}^{n+1} (-1)^{n-k+1} s(n+1,k)2^k \left( \frac{2^{k+1}}{k+1}+(-1)^{k}\frac{(n+2)^{k+1}}{k+1}\right).
    \end{split}
\end{equation}
Note that the second term in the above expression,
\begin{equation}
    \frac{1}{(n+4)(n+1)!} \sum_{k=0}^{n+1} (-1)^{n+1} s(n+1,k)2^k\frac{(n+2)^{k+1}}{k+1}
\end{equation}
is always manifestly positive, since for non-zero $a_{n,0}$, $n$ must be odd. It remains to show the positivity of the first term to ensure $a_{n,0}>0$.

\section{In general dimensions}
In this section, we will write the pseudo-generating function  $g_{n,k}^{\alpha}(h)$ in general dimensions (remember that $\alpha=(D-3)/2$, see (\ref{alpha})). Introducing the generating functions for $a_{n,l}$ from (\ref{a_{n,l}})
\begin{equation}\label{Generating_in_D}
   G^{\alpha}_{n}(h)=\sum_{j=0}^{\infty} \Lambda(j,\alpha) \frac{a_{n,j}}{h^{j+1}}=\frac{1}{(n+4)(n+1)!}\left[\frac{4}{(n+4)^2} \right]^{\alpha-\frac{1}{2}} \int_{0}^{n+4} dt \frac{(-t+2)_{n+1}[t(n+4-t)]^{\alpha-\frac{1}{2}}}{[(h-1)^2+\frac{4ht}{n+4}]^{\alpha}} ,
\end{equation}
where we have used the standard generating function for $C_{j}^{(\alpha)}$:
\begin{equation}
    \sum_{j=0}^{\infty} C_{j}^{(\alpha)}(x) t^j = [1-2xt+t^2]^{-\alpha}.
\end{equation}
We can now use (\ref{series_expansion}) and the following series expansions
\begin{equation}
    \begin{split}
    (n+4-t)^{\alpha-\frac{1}{2}}=\sum_{p=0}^{\infty} \binom{\alpha-\frac{1}{2}}{p}(-1)^{p}(n+4)^{\alpha-\frac{1}{2}-p}t^p,
    \end{split}
\end{equation}
to rewrite the generating function (\ref{Generating_in_D}) as
\begin{equation}\label{extra_p}
    \begin{split}
         G^{\alpha}_{n}(h)=\frac{1}{(n+4)(n+1)!}\left[\frac{4}{(n+4)^2} \right]^{\alpha-\frac{1}{2}}\sum_{k=0}^{n+1}(-1)^{n+1-k}s(n+1,k) \sum_{p=0}^{\infty} \binom{\alpha-\frac{1}{2}}{p}(-1)^{p}(n+4)^{\alpha-\frac{1}{2}-p} \, g^{\alpha}_{n,k,p}(h),
    \end{split}
\end{equation}
with the `pseudo-generating function' $g^{\alpha}_{n,k,p}(h)$ defined as  \begin{equation}
    g^{\alpha}_{n,k,p}(h)=\int_{0}^{n+4}dt\,\frac{(-t+2)^k \,t^{p+\alpha-\frac{1}{2}}}{[(h-1)^2+\frac{4ht}{n+4}]^{\alpha}}.
\end{equation}
We can evaluate it, as before, in terms of Appell hypergeometric function of two variables
\begin{equation}\label{general_dim_g}
     g^{\alpha}_{n,k,p}(h)=\frac{2^{k+1}}{2\alpha+2p+1}\frac{(n+4)^{p+\alpha+\frac{1}{2}}}{(h-1)^{2\alpha}} \text{Appell}F_{1}\left( p+\alpha+\frac{1}{2};-k,\alpha;p+\alpha+\frac{3}{2};\frac{n+4}{2},-\frac{4h}{(h-1)^2}\right).
\end{equation}
 Note the extra $p$-sum in (\ref{extra_p}) compared to $\alpha=1/2$ case, which complicates the calculation in general dimensions. We might have to find the coefficient with minimum power of $h$ in the expansion of the above expression (\ref{general_dim_g}) to hope for gettring $a_{n,n+1}$. For that, a representation of this Appell function (\ref{general_dim_g}) like (\ref{g[n,k]1}) would be highly desirable.  
 \par 
 In fact we can calculate $a_{n,l}$ numerically for the first few levels and check that the positivity condition breaks down for $D>26$ \cite{Caron-Huot:2016icg} i.e, above the critical dimension for bosonic string theory!

\end{document}